\documentclass[fleqn,usenatbib]{mnras}

\usepackage{newtxtext,newtxmath}

\usepackage[T1]{fontenc}

\DeclareRobustCommand{\VAN}[3]{#2}
\let\VANthebibliography\thebibliography
\def\thebibliography{\DeclareRobustCommand{\VAN}[3]{##3}\VANthebibliography}

\usepackage{graphicx}	

\usepackage{cleveref}
\crefname{figure}{Fig.}{Figs.}%
\crefname{equation}{equation}{equations}

\newcommand{\T}[1]{T_{\mathrm{#1}}}
\newcommand{\psd}[1]{P_{\mathrm{#1}}}
\newcommand{\G}[1]{\Gamma_{\mathrm{#1}}}
\newcommand{\Ga}{\Gamma_{\mathrm{cal}}}
\newcommand{\Gr}{\Gamma_{\mathrm{rec}}}
\newcommand{\y}{\mathbfit{T}_{\mathrm{cal}}}
\newcommand\given[1][]{\:#1\vert\:}

\graphicspath{ {./images/} }

\title[Bayesian calibration for 21-cm experiments]{Bayesian noise wave calibration for 21-cm global experiments}

\author[I. L. V. Roque et al.]{
I. L. V. Roque,$^{1}$\thanks{E-mail: ilvr2@cam.ac.uk}
W. J. Handley$^{1,2}$
and N. Razavi-Ghods$^{1}$
\\
$^{1}$Astrophysics Group, Cavendish Laboratory, JJ Thomson Avenue, Cambridge, CB3 0HE, UK\\
$^{2}$Kavli Institute for Cosmology, Madingley Road, Cambridge, CB3 0HA, UK\\
}

\date{Accepted XXX. Received YYY; in original form ZZZ}

\pubyear{2020}

\begin{document}
\label{firstpage}
\pagerange{\pageref{firstpage}--\pageref{lastpage}}
\maketitle

\begin{abstract}
Detection of millikelvin-level signals from the ‘Cosmic Dawn’ requires an unprecedented level of sensitivity and systematic calibration. We report the theory behind a novel calibration algorithm developed from the formalism introduced by the EDGES collaboration for use in 21-cm experiments. Improvements over previous approaches are provided through the incorporation of a Bayesian framework and machine learning techniques such as the use of Bayesian evidence to determine the level of frequency variation of calibration parameters that is supported by the data, the consideration of correlation between calibration parameters when determining their values and the use of a conjugate-prior based approach that results in a fast algorithm for application in the field. In self-consistency tests using empirical data models of varying complexity, our methodology is used to calibrate a 50 $\Omega$ ambient-temperature load. The RMS error between the calibration solution and the measured temperature of the load is 8 mK, well within the $1\sigma$ noise level. Whilst the methods described here are more applicable to global 21-cm experiments, they can easily be adapted and applied to other applications, including telescopes such as HERA and the SKA.
\end{abstract}

\begin{keywords}
instrumentation: interferometers -- methods: statistical -- dark ages, reionization, first stars
\end{keywords}



\section{Introduction}\label{intro}
For nearly a century, scientists have been using radio-frequency instruments to advance the study of astronomy and complement information from the visual regime of the electromagnetic spectrum \citep{21in21}. As we begin to take measurements of the early universe, these instruments must continue to evolve to support observations. Unexplored cosmic information from the Epoch of Reionisation and Cosmic Dawn redshifted into the radio spectrum could provide constraints on fundamental physics such as primordial black holes, galaxy formation, and universal curvature as discussed in \citet{furAst}. A unique probe of phenomena from the early cosmos is the hydrogen that inundates the intergalactic medium (IGM). Heating and cooling of the IGM associated with hydrogen's absorption and emission of 21-cm photons produce a dynamic brightness temperature relative to the cosmic microwave background temperature, tracing the evolution of surrounding structure during the Cosmic Dawn. The brightness temperature of this 21-cm photon signal can be described by
\begin{equation}
    \label{brightnessTemp}
    \begin{aligned}
    T_{21}(z) \approx & \ 0.023 \mathrm{K} \ \times \\
    & x_{\ion{H}{i}}(z) \left[ \left(\frac{0.15}{\Omega_{\mathrm{m}}} \right)\left(\frac{1+z}{10}\right) \right]^{\frac{1}{2}} \left(\frac{\Omega_{\mathrm{b}}h}{0.02}\right)\left[1-\frac{T_{\mathrm{R}}(z)}{T_{\mathrm{S}}(z)}\right],
    \end{aligned}
\end{equation}
which is heavily dependent on environmental factors of the early universe such as $x_{\ion{H}{i}}$, the fraction of neutral hydrogen, $\Omega_{\mathrm{m}}$ and $\Omega_{\mathrm{b}}$, the matter and baryon densities with respect to the universal critical density for a flat universe and Hubble's constant. Here, the $0.023$ is a constant from atomic-line physics. $T_{\mathrm{R}}$ is the background radiation temperature and $T_{\mathrm{S}}$ is known as the `21-cm spin temperature', which is related to the kinetic temperature of neutral hydrogen gas in the IGM \citep{radiationTemp, spinTemp}. This cosmic hydrogen signature measurable in the spectral sky has been redshifted to wavelengths under 200 MHz through the expansion of the universe as discussed in \citet{21in21}.

There has been a recent surge in the field of 21-cm cosmology following the reported detection of an absorption feature consistent with a Cosmic Dawn signature.  This was reported by the Experiment to Detect the Global EoR Signature (EDGES) in early 2018 from measurements of a sky-averaged radio spectrum \citep{monsalve}. The signal, centred at 78 MHz with a width corresponding to a period between 180 million and 270 million years after the Big Bang, matches the theoretical position in frequency, but its depth of $\sim 0.5$ K is a factor of two greater than the largest predictions from theoretical models \citep{fialkov}. This discrepancy would suggest that the temperature difference between the IGM and the cosmic microwave background was much larger than previously thought and would require new physics to explain, such as dark matter-baryon interactions \citep{darkmatter} or excess radio backgrounds \citep{radio}. 

Another possible explanation for this discrepancy is that the measured signal is not cosmological but of systematic origin. This may be the case in EDGES due to some of the methodology used, such as a potentially unphysical foreground removal method and calibration of the receiver in a separate environment from the data acquisition \citep{hills, nimaRise}. In this paper, we present a novel calibration algorithm that improves on the work of the EDGES team \citep{rogers} through the utilisation of a Bayesian framework to promote efficient use of the data to remove systematics. Using conjugate priors and machine learning techniques, our pipeline can be applied in the field with the collection of data with additional capabilities for optimising individual noise wave parameters and incorporating correlations between them.

This paper is organised as follows. In \cref{theory} we review the methodology behind calibration using noise waves as well as present a Bayesian framework that provides greater flexibility in radiometer calibration. \Cref{mockdata} describes the process of using mock data sets modelled after empirical measurements of reflection coefficients with the incorporation of a realistic noise model to evaluate our pipeline.


\section{Methods}\label{theory} 
In this section, we detail the methodology behind radiometer calibration using noise wave parameters. An overview of global signal measurement are outlined in \cref{measSig}. \Cref{edgesCalibration} summarises the basic procedure with some mathematical improvements while \cref{chap:bayes} describes our Bayesian framework and its associated advantages.

\subsection{Measuring the global signal}\label{measSig}
\begin{figure*}
    \centering
    \includegraphics[scale=0.35]{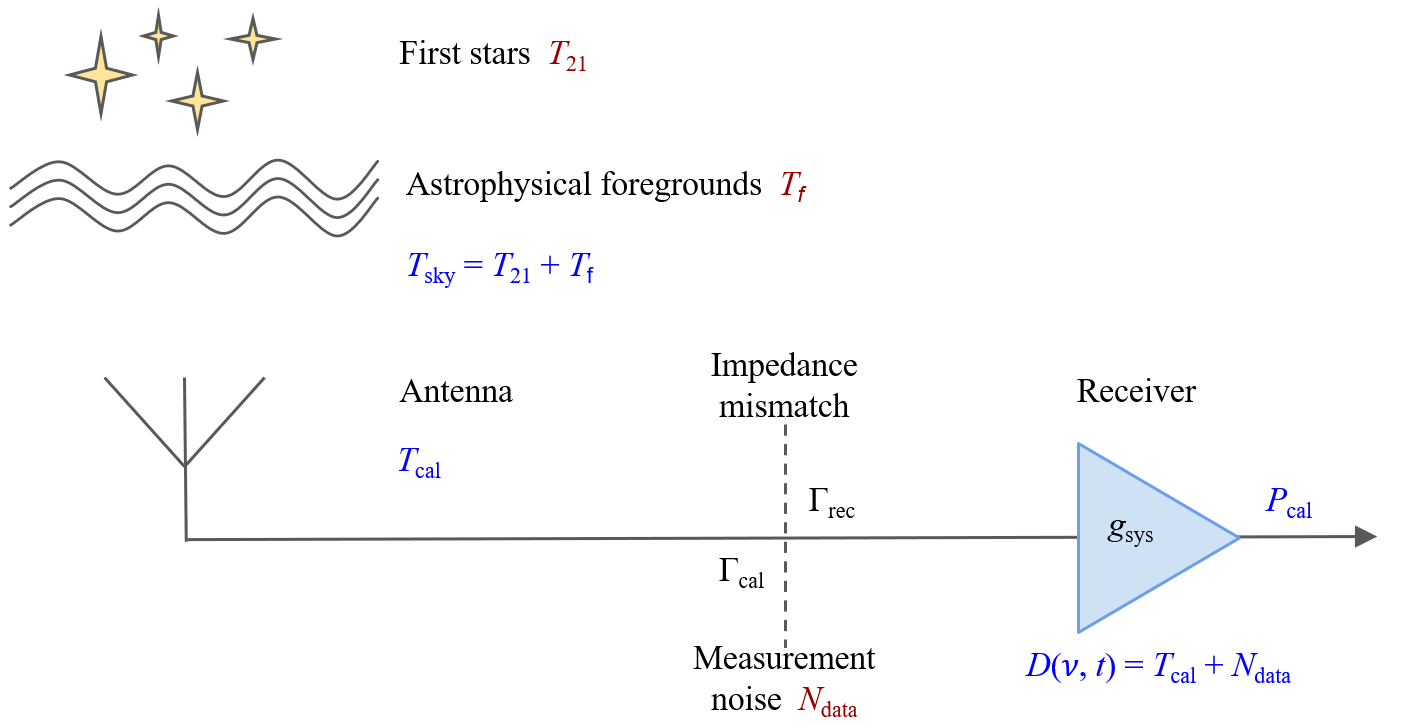}
    \caption{Diagram showing the evolution of the 21-cm signal hampered by astrophysical foregrounds, convolvution with the antenna beam and the emergence of measurement noise before calibration to retrieve the antenna temperature.}
    \label{fig:nsfig}
\end{figure*}

The noise necessitating calibration emerges during measurement-taking. In an averaged or global experiment, the sky temperature \mbox{$\T{sky}(\Omega, \nu, t)$} is a function of the direction $\Omega$, frequency $\nu$ and time $t$. This can be broken down into two primary components: the global 21-cm signal $T_{21}$ and astrophysical foregrounds $\T{f}$
\begin{equation}
    \label{tsky}
    \T{sky}(\Omega, \nu, t) = T_{21}(\nu) + \T{f}(\Omega, \nu, t).
\end{equation}
The antenna measures the sky signal convolved with the normalised antenna directivity $B$. The process of measurement introduces the random noise term $N_{\mathrm{data}}$.
\begin{equation}\label{bayestsource}
    D(\nu, t) = \int \T{sky}(\Omega, \nu, t) B(\Omega, \nu)\mathrm{d}\Omega + N_{\mathrm{data}}.
\end{equation}
Our 21-cm signature can thus be represented as
\begin{equation}\label{signal}
  T_{21} \approx D(\nu, t) - \int\T{f}(\Omega, \nu, t)B(\Omega, \nu)\mathrm{d}\Omega - N_{\mathrm{data}}.
\end{equation}
Here, the integral is assessed through foreground and beam modelling techniques such as those discussed in \citet{dom} while modelling of $N_{\mathrm{data}}$ from the statistical properties of $D(\nu, t)$ is accomplished by a calibration algorithm as articulated in this paper and outlined in \cref{fig:nsfig}. Having a fully Bayesian framework when modelling the beam, the sky and the systematics has major advantages for global 21-cm experiments such as REACH \citep{reach}, as it provides the greatest flexibility in being able to model all effects and jointly fit for them.

\subsection{Calibration methodology}\label{edgesCalibration}
\begin{figure}
  \centering
  \includegraphics[width=\columnwidth]{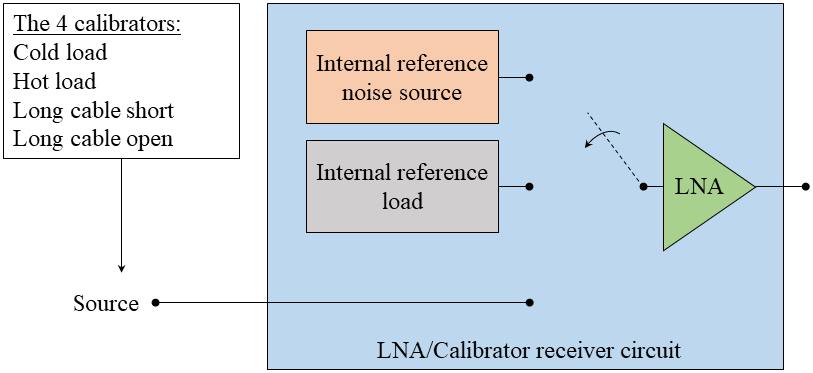}
  \caption{Diagram of a typical calibration setup. For characterisation of the receiver, a switch cycles between a calibrator connected to the input and the two internal references. \label{f:dickeswitchpic}}
\end{figure}

The standard calibration strategy follows the method introduced by Dicke to characterise systematic features in radio frequency instruments \citep{dickeplus} and is widely used in experiments such as EDGES \citep{calpap} and LOFAR \citep{lofarCal} to evaluate the spectral index of the sky's diffuse radio background \citep{rogers}. This technique involves measurements of two internal reference standards; a load and a noise source, in addition to a series of external calibration sources attached to the receiver input in lieu of the antenna. These include an ambient-temperature ‘cold’ load, a ‘hot’ load heated to $\sim 400$ K, an open-ended cable and a shorted cable. A block diagram showing this arrangement is presented in \cref{f:dickeswitchpic}. 

When calibrating the receiver, reflection coefficients are taken of the calibration source connected to the receiver input ($\Ga$) and of the receiver itself ($\G{rec}$) as well as power spectral densities (PSDs) of the input ($\psd{cal}$), the internal reference load ($\psd{L}$) and the internal reference noise source ($\psd{NS}$) \citep{calpap}. These measurements are used to calculate a preliminary `uncalibrated' antenna temperature $\T{cal}^*$
\begin{equation}
  \label{eqn:tantstar}
  \T{cal}^* = \T{NS} \left(\frac{\psd{cal}-\psd{L}}{\psd{NS}-\psd{L}}\right) + \T{L},
\end{equation}
where $\T{L}$ and $\T{NS}$ are assumptions for the noise temperature of the internal reference load and excess noise temperature of the internal noise source above ambient, respectively. This initial calculation is used to calibrate out any time-dependent system gain that emerges from a series of filters, amplifiers and cables, as well as the analogue-to-digital converter within the experimental apparatus \citep{calpap}. Each PSD measurement can be expressed in terms of specific response contributions as detailed in \citet{monsalve}
\begin{equation}
  \label{eqn:pant}
  \begin{aligned}
  \psd{cal} = g_{\mathrm{sys}} \Bigg[ &\T{cal}\left(1-|\Ga|^2\right)\left|\frac{\sqrt{1 - |\G{rec}|^2}}{1-\Ga\G{rec}}\right|^2 \\
  + & \T{unc}|\Ga|^2\left|\frac{\sqrt{1 - |\G{rec}|^2}}{1-\Ga\G{rec}}\right|^2 \\
  + & \T{cos}\operatorname{Re}\left(\Ga\frac{\sqrt{1 - |\G{rec}|^2}}{1-\Ga\G{rec}}\right) \\
  + & \T{sin}\operatorname{Im}\left(\Ga\frac{\sqrt{1 - |\G{rec}|^2}}{1-\Ga\G{rec}}\right) 
  + T_0 \Bigg].
  \end{aligned}
\end{equation}

Here, $g_{\mathrm{sys}}$ is the system gain referenced to the receiver input and $\T{cal}$ is our calibrated input temperature. $\T{unc}$, $\T{cos}$, and $\T{sin}$ are the ‘noise wave parameters’ introduced by \citet{Meys} to calibrate the instrument. $\T{unc}$ represents the portion of noise reflected by the antenna that is uncorrelated with the output noise of the low noise amplifier (LNA). $\T{cos}$ and $\T{sin}$ are the portions of reflected noise correlated with noise from the LNA \citep{calpap, rogers}. In the EDGES experiment, these calibration quantities are modelled using seven-term polynomials in frequency.

The PSDs for the internal reference load and noise source can similarly be expressed as in \cref{eqn:pant}. However, since the reflection coefficients of the internal references are typically less than 0.005, they are taken to be zero in order to simplify the equations
\begin{equation}
  \label{eqn:pl}
  \psd{L} = g_{\mathrm{sys}}^*[\T{L}\left(1-|\G{rec}|^2\right)+T_{0}^*],
\end{equation}
\begin{equation}
  \label{eqn:pns}
  \psd{NS} = g_{\mathrm{sys}}^*[\left(\T{L}+\T{NS}\right)\left(1-|\G{rec}|^2\right)+T_{0}^*].
\end{equation}

As shown in \cref{f:dickeswitchpic}, the internal references may be on a separate reference plane than the receiver input, resulting in a system gain $g_{\mathrm{sys}}^*$ and a noise offset $T_{0}^*$ different from those defined in \cref{eqn:pant}. This effect is taken into account by two additional scale and offset parameters, $C_1$ and $C_2$, introduced by EDGES \citep{calpap}. 

Since $C_1$ and $C_2$ also correct for first-order assumptions in the noise temperatures of the internal reference load and noise source, we have chosen to absorb these terms into $\T{L}$ and $\T{NS}$. This adjustment allows all calibration parameters, $\T{unc}$, $\T{cos}$, $\T{sin}$, and an ‘effective’ $\T{NS}$ and $\T{L}$, to be solved for in units of kelvin, facilitating a joint solution of parameters. Expanding \cref{eqn:tantstar} using \cref{eqn:pant,eqn:pl,eqn:pns} yields a linear identity providing a relationship between the uncalibrated input temperature and a final calibrated temperature of any device connected to the receiver input
\begin{equation}
  \label{eqn:caleqn}
  \begin{aligned}
  \T{NS}\left( \frac{\psd{cal} - \psd{L}}{\psd{NS} - \psd{L}} \right) + \T{L}&= \T{cal}\left[ \frac{1-|\Ga|^2}{|1-\Ga\G{rec}|^2} \right] \\
  & + \T{unc}\left[ \frac{|\Ga|^2}{|1-\Ga\G{rec}|^2} \right] \\
  & + \T{cos}\left[ \frac{\operatorname{Re}\left(\frac{\Ga}{1-\Ga\G{rec}}\right)}{\sqrt{1-|\G{rec}|^2}} \right] \\
  & + \T{sin}\left[ \frac{\operatorname{Im}\left(\frac{\Ga}{1-\Ga\G{rec}}\right)}{\sqrt{1-|\G{rec}|^2}} \right], \\ 
  \end{aligned}
\end{equation}
where all parameters are frequency-dependent. This is not explicitly shown for simplicity of notation. For estimation of the noise wave parameters, $\T{cal}$, $\Ga$ and $\G{rec}$ are measured along with the PSDs while $g_{\mathrm{sys}}$ and $\T{0}$ are calibrated out. The cold and hot loads exhibit the main temperature references needed for $\T{L}$ and $\T{NS}$. The cables facilitate the derivation of the noise wave parameters describing spectral ripples from the noise properties of the receiver by acting as antennas looking at an isotropic sky with temperatures equal to the cables' physical temperatures \citep{rogers}.


\subsection{Bayesian calibration framework}\label{chap:bayes}
One possible source of systematics in the calibration methodology used by EDGES comes from measuring the response of the four external calibrators along with the receiver reflection coefficient in a laboratory away from where the instrument is actually deployed \citep{monsalve}.  This process, especially with regards to how calibration parameters change, can be non-trivial. Furthermore, the fixed polynomial order used by EDGES for all noise wave parameters may underfit or overfit individual parameters and thus `fit out' data useful for determining systematics or potentially even the 21-cm signal itself if a joint fit is performed. 

\begin{figure*}
    \centering
    \includegraphics[width=\textwidth]{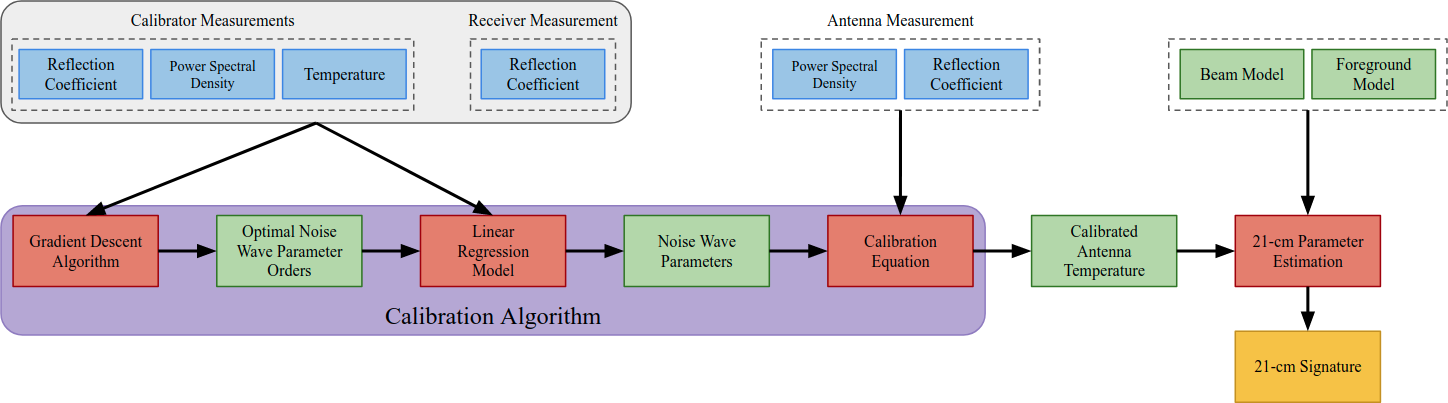}
    \caption{Outline of the Bayesian calibration algorithm. Blue blocks represent data to be taken, red blocks represent calculations and green blocks represent calculation outputs.}
    \label{flowchart}
\end{figure*}

In response to these issues, we have developed a calibration pipeline that improves on the strategies presented in \cref{edgesCalibration}. We introduce a novel Bayesian methodology using conjugate priors for a dynamic application of our algorithm to be run with data collection regardless of system complexity. Also included are model selection methods using machine learning techniques for the optimisation of individual noise wave parameters to combat overfitting and underfitting, the results of which converge with that of a least-squares approach when wide priors are adopted. Our pipeline easily incorporates many more calibrators than the standard four shown in \cref{f:dickeswitchpic} to increase constraints on noise wave parameters while identifying possible correlations between them. A schematic of the improved calibration method is shown in \cref{flowchart}.

In order to simplify our calibration approach, we first define the following terms
\begin{equation}
  X_{\mathrm{unc}} = -\frac{|\Ga|^2}{ 1-|\Ga|^2}, 
\end{equation}
\begin{equation}\label{eqn:xl}
  X_{\mathrm{L}} = \frac{|1-\Ga\Gr|^2}{1-|\Ga|^2},
\end{equation}
\begin{equation}
  X_{\mathrm{cos}} = -\operatorname{Re}\left(\frac{\Ga}{1-\Ga\Gr} \times \frac{X_{\mathrm{L}}}{\sqrt{1-|\Gr|^2}}\right),
\end{equation}
\begin{equation}
  X_{\mathrm{sin}} = -\operatorname{Im}\left(\frac{\Ga}{1-\Ga\Gr} \times \frac{X_{\mathrm{L}}}{\sqrt{1-|\Gr|^2}}\right),
\end{equation}
\begin{equation}\label{eqn:xns}
  X_{\mathrm{NS}} = \left( \frac{P_{\mathrm{cal}}-P_{\mathrm{L}}}{P_{\mathrm{NS}}-P_{\mathrm{L}}} \right) X_{\mathrm{L}},
\end{equation}
which represent initial calibration measurements on $D$ in the frequency domain for the characterisation of $N_{\mathrm{data}}$ from \cref{bayestsource} via our noise wave parameters. It is expected that calibration-related deviations of $D$ in the time domain are sufficiently curtailed through practical strategies such as temperature control of the receiver environment. Incorporating these into \cref{eqn:caleqn}, with some rearrangement, then gives the equation
\begin{equation}
  X_{\mathrm{unc}}\T{unc} + X_{\mathrm{cos}}\T{cos} + X_{\mathrm{sin}}\T{sin} + X_{\mathrm{NS}}\T{NS} + X_{\mathrm{L}}\T{L} = \T{cal},
\end{equation}
at each frequency. Here, there are no squared or higher-order terms, allowing us to take advantage of the linear form by grouping the data and noise wave parameters into separate matrices
\begin{align}\label{eqn:theta}
  \mathbfss{X} &\equiv \begin{pmatrix} 
  X_\mathrm{unc} \quad 
  X_\mathrm{cos} \quad
  X_\mathrm{sin} \quad
  X_\mathrm{NS} \quad
  X_\mathrm{L} \end{pmatrix},\nonumber\\
  \boldsymbol{\Theta} &\equiv \begin{pmatrix} 
  T_\mathrm{unc}\quad
  T_\mathrm{cos}\quad
  T_\mathrm{sin}\quad
  T_\mathrm{NS}\quad
  T_\mathrm{L}\end{pmatrix}^\top.
\end{align}

In these equations, all of our data; the reflection coefficient measurements and power spectral densities, are grouped in an $\mathbfss{X}$ vector which forms a matrix where one of the axes is frequency. The calibration parameters as frequency-dependent polynomials of varying degree are collected into a $\boldsymbol{\boldsymbol{\Theta}}$ vector which serves as our model describing $N_{\mathrm{data}}$. Applying these definitions condenses the calibration equation into
\begin{equation}\label{eqn:linearmodel}
  \y = \mathbfss{X}\boldsymbol{\boldsymbol{\Theta}}+\sigma,
\end{equation}
where $\y$ is a vector over frequency and $\sigma$ is a noise vector representing our error. Since EDGES assumes that each power spectral density measurement is frequency independent, we have assumed that $\sigma$ is a multivariate normal distribution. This assumption is implicit in the EDGES analysis in which they use a least-squares minimisation approach for solving model parameters. 

For calibration of the receiver, we are concerned with the construction of predictive models of the noise wave parameters, $\boldsymbol{\Theta}$, in the context of some dataset, $\mathbfit{T}$. We can use $\boldsymbol{\Theta}$ to calculate the probability of observing the data given a specific set of noise wave parameters:
\begin{equation}\label{likelihood}
  \begin{aligned}
    p\big(\mathbfit{T} \given[\big] \boldsymbol{\Theta}, \sigma^2\big) &= \\ & \frac{1}{2\pi \sigma^2}^{N/2}\exp{ \Bigg\{ -\frac{1}{2\sigma^2}\left(\mathbfit{T}-\mathbfss{X}\boldsymbol{\Theta}\right)^{\top}\left(\mathbfit{T} -\mathbfss{X}\boldsymbol{\Theta}\right) \Bigg\}},
  \end{aligned}
\end{equation}
where, $N$ is the number of measurements. This distribution on the data is the \textit{likelihood}. For the purposes of calibration, $\mathbfit{T}$ may be $\y$ measurements or alternatively, $\mathbfit{T}_{\mathrm{sky}}$ for prediction of a sky signal. Our model must also specify a \textit{prior} distribution, quantifying our initial assumptions on the values and spread of our noise wave parameters which we specify as a multivariate normal inverse gamma distribution:
\begin{equation}
  \begin{aligned}
  \label{eqn:prior}
  p\left(\boldsymbol{\Theta}, \sigma^2\right) \propto & \left(\frac{1}{\sigma^2}\right)^{a+1+\left(d/2\right)} \times \\ &\exp \left[ -\frac{1}{\sigma^2}\{b+\frac{1}{2}\left(\boldsymbol{\Theta}-\boldsymbol{\mu}_{\boldsymbol{\Theta}}\right)^{\top}\mathbfss{V}_{\boldsymbol{\Theta}}^{-1}\left(\boldsymbol{\Theta}-\boldsymbol{\mu}_{\boldsymbol{\Theta}}\right)\} \right],
  \end{aligned}
\end{equation}
which is proportional up to an integration constant. Here, $a$ and $b$, which are greater than zero, along with $\mathbfss{V}_{\boldsymbol{\Theta}}$ and $\boldsymbol{\mu}_{\boldsymbol{\Theta}}$ represent our prior knowledge on the noise wave parameters. $d$ is the length of our vector $\boldsymbol{\Theta}$.

\Cref{likelihood} is determined by a set of values for our model $\boldsymbol{\Theta}$. We can marginalise out the dependence on $\boldsymbol{\Theta}$ and our noise term by integrating over the prior distribution by both $\boldsymbol{\Theta}$ and $\sigma^2$ at once. Following the steps in \citet{banerjee}
\begin{equation}
    \begin{aligned} \label{eqn:ev}
    p\left(\y\right) &= \int p\left(\y \given[\big] \boldsymbol{\Theta}, \sigma^2\right) p\left(\boldsymbol{\Theta}, \sigma^2\right) \mathrm{d}\boldsymbol{\Theta} \mathrm{d}\sigma^2\\
    &= \frac{b^a\Gamma\left(a^*\right)\sqrt{|\mathbfss{V}^*|}}{{b^*}^{a^*}\Gamma\left(a\right)\sqrt{|\mathbfss{V}_{\boldsymbol{\Theta}}|}}(2\pi)^{-N/2}, \\
    \end{aligned}
\end{equation}    
where 
\begin{equation}\label{starred}
    \begin{aligned}   
    a^* &= a + \frac{N}{2}, \\
    b^* &= b + \frac{1}{2}[\boldsymbol{\mu}_{\boldsymbol{\Theta}}^{\top}\mathbfss{V}_{\boldsymbol{\Theta}}^{-1}\boldsymbol{\mu}_{\boldsymbol{\Theta}} + \y^{\top}\y - \boldsymbol{\mu}^{*\top}\mathbfss{V}^{*-1}\boldsymbol{\mu}^*], \\
    \boldsymbol{\mu}^* &= \left(\mathbfss{V}_{\boldsymbol{\Theta}}^{-1} + \mathbfss{X}^{\top}\mathbfss{X}\right)^{-1}\left(\mathbfss{V}_{\boldsymbol{\Theta}}^{-1}\boldsymbol{\mu}_{\boldsymbol{\Theta}} + \mathbfss{X}^{\top}\y\right), \\
    \mathbfss{V}^* &= \left(\mathbfss{V}_{\boldsymbol{\Theta}}^{-1} + \mathbfss{X}^{\top}\mathbfss{X}\right)^{-1}, \\
    \end{aligned}
\end{equation}
and $\Gamma\left(x\right)$ represents the Gamma function, not to be confused with the notation for our reflection coefficients. \Cref{eqn:ev} is the \textit{evidence}, which gives the probability of observing the data $\y$ given our model.\footnote{It is in fact better to use the equivalent more numerically stable expression
$b^*=b + \boldsymbol{q}^{\top} \boldsymbol{q} + \boldsymbol{q}^{\top} \mathbfss{X} \mathbfss{V}_{\boldsymbol{\Theta}} \mathbfss{X}^{\top} \boldsymbol{q}$,  where $\boldsymbol{q}= \y-\mathbfss{X}\boldsymbol{\mu}^*$ to avoid cancellation of large terms.}  

With the prior distribution specified, we use Bayes' equation to invert the conditioning of the likelihood and find the \textit{posterior} using the likelihood, prior and evidence:
\begin{equation}
    p\left(\boldsymbol{\Theta}, \sigma^2 \given[\big] \y\right) = \frac{p\left(\y \given[\big]  \boldsymbol{\Theta}, \sigma^2\right)p\left(\boldsymbol{\Theta}, \sigma^2\right)}{p\left(\y\right)}.
\end{equation}
Similarly from \citet{banerjee}, this can be written as
\begin{equation}
    \begin{aligned} \label{eqn:post}
    p\Bigl(\boldsymbol{\Theta},\sigma^2 \given[\big] & \y\Bigl) \propto \left(\frac{1}{\sigma^2}\right)^{a^* + \frac{d}{2} + 1} \times \\ 
    & \exp{ \Bigg\{ -\frac{1}{\sigma^2} \Bigg[ b^* + \frac{1}{2}\left(\boldsymbol{\Theta} - \boldsymbol{\mu}^*\right)^{\top}\mathbfss{V}^{*-1}\left(\boldsymbol{\Theta} - \boldsymbol{\mu}^*\right) \Bigg] \Bigg\} }.
    \end{aligned}
\end{equation}

The posterior distribution represents the uncertainty of our parameters after analysis, reflecting the increase in information \citep{nagel}. We highlight the difference between the `likelihood-only' least-squares approach versus the Bayesian approach with the former being a special case of the latter with very wide priors demonstrable when $\mathbfss{V}_{\boldsymbol{\Theta}} \rightarrow \infty \Rightarrow \mathbfss{V}_{\boldsymbol{\Theta}}^{-1} \rightarrow 0$, and $\boldsymbol{\mu}^*$ becomes $\boldsymbol{\Theta}$. The transition from `non-starred' variables to `starred' variables represents our `Bayesian update' of the prior to the posterior noise wave parameters in light of the calibration data $\y$.

As we can see, the posterior distribution is in the same probability distribution family as \cref{eqn:prior}, making our prior a \textit{conjugate prior} on the likelihood distribution. The use of conjugate priors gives a closed-form solution for the posterior distribution through updates of the prior hyperparameters via the likelihood function \citep{banerjee, orloff}. The resulting numerical computation is many orders of magnitude faster than MCMC methods relying on full numerical sampling and permits an in-place calculation in the same environment as the data acquisition. This becomes particularly useful for the speed of the algorithm as frequency dependence is introduced in which the computations would not be manageable without conjugate gradients. 

To allow for a smooth frequency dependency, we promote each of our noise wave parameters in \cref{eqn:theta} to a vector of polynomial coefficients
\begin{equation}
    \T{i} = \begin{pmatrix}
    \T{i}^{[0]}, & \T{i}^{[1]}, & \T{i}^{[2]}, & ..., & \T{i}^{[n]}
    \end{pmatrix},
\end{equation}
where $i$ is our noise wave parameter label; $i \in \{\mathrm{unc, \ cos, \ sin , \ NS, \ L}\}$, modelled using $n+1$ polynomial coefficients. Likewise
\begin{equation}
    \mathbfss{X}_{i} = \begin{pmatrix}
    \mathbfss{X}_{i}, & \mathbfss{X}_{i}\left(\frac{\nu}{\nu_0}\right), & \mathbfss{X}_{i}{\left(\frac{\nu}{\nu_0}\right)}^2, & ..., &  \mathbfss{X}_{i}{\left(\frac{\nu}{\nu_0}\right)}^{n}
    \end{pmatrix},
\end{equation}
where $\nu$ is a vector of input frequencies which are raised to powers up to $n$. For a vector of $n$'s attributed to our calibration parameters, under this notation multiplication in \cref{eqn:linearmodel} is element-wise and \cref{eqn:ev} is effectively $p\left(\y|\mathbfit{n}\right)$. Assuming a uniform prior on $\mathbfit{n}$, inverting Bayes' theorem gives $p\left(\mathbfit{n}|\y\right)$ for use in model comparison in which the relative probabilities of models can be evaluated in light of the data and priors. Occam’s razor advises whether the extra complexity of a model is needed to describe the data \citep{trotta}, permitting optimisation of the polynomial orders for individual noise wave parameters as detailed in \cref{chap:opt}. By taking a random sampling of the resulting posterior, we characterise the noise wave parameters as multivariate distributions depicted in contour plots which exhibit a peak value accompanied by $1\sigma$ and $2\sigma$ variance as well as correlation between parameters inferred from a covariance matrix.

Following characterisation of the receiver, we next apply the $\y$ from our calibration to a set of raw antenna data $\hat{\mathbfss{X}}$ for prediction of our sky signal, $\mathbfit{T}_{\mathrm{sky}}$, from \cref{bayestsource}. The predictions for the data follow from the \emph{posterior predictive distribution}
\begin{equation}
    p\left(\mathbfit{T}_{\mathrm{sky}} \given[\big] \mathbfit{T}_{\mathrm{cal}} \right) = \int p\left( \mathbfit{T}_{\mathrm{sky}} \given[\big] \boldsymbol{\Theta},\sigma^2 \right) p \left( \boldsymbol{\Theta},\sigma^2 \given[\big] \mathbfit{T}_{\mathrm{cal}} \right) \mathrm{d}\boldsymbol{\Theta}\mathrm{d}\sigma^2.
\end{equation}
The first probability in the integral is the likelihood for our antenna measurement $\mathbfit{T}_{\mathrm{sky}}$ and the second is our posterior from \cref{eqn:post}. Following the steps in \citet{banerjee}, this can be shown to be a multivariate Student's t-distribution written as:
\begin{equation}\label{predictive}
    \begin{aligned}
    p\Big( & \mathbfit{T}_{\mathrm{sky}} \given[\big] \mathbfit{T}_{\mathrm{cal}} \Big) = \frac{\Gamma\left( a^* + \frac{d}{2} \right)}{\Gamma\left( a^* \right)\pi^{\frac{d}{2}}|2b^*\left( I + \hat{\mathbfss{X}}\mathbfss{V}^*\hat{\mathbfss{X}}^{\top}\right)|^{\frac{1}{2}}}
    \\ & \times
    \left[ 1 + \frac{\left( \mathbfit{T}_{\mathrm{sky}} - \hat{\mathbfss{X}}\boldsymbol{\mu}^* \right)^{\top} \left( I + \hat{\mathbfss{X}}\mathbfss{V}^*\hat{\mathbfss{X}}^{\top} \right)^{-1} \left( \mathbfit{T}_{\mathrm{sky}} - \hat{\mathbfss{X}}\boldsymbol{\mu}^* \right)}{2b^*} \right]^{-\left( a^* + \frac{d}{2} \right)},
    \end{aligned}
\end{equation}
where $I$ is the $N \times N$ identity matrix and $a^*$, $b^*$, $\boldsymbol{\mu}^*$ and $\mathbfss{V}^*$ are defined in \cref{starred}. This new distribution on $\mathbfit{T}_{\mathrm{sky}}$ corresponds to a set of points with error bars and represents the calibrated sky temperature as the output of the receiver.


\section{Empirical modelling and simulations}\label{mockdata} 
\begin{figure}
  \centering
  \includegraphics[width=\columnwidth]{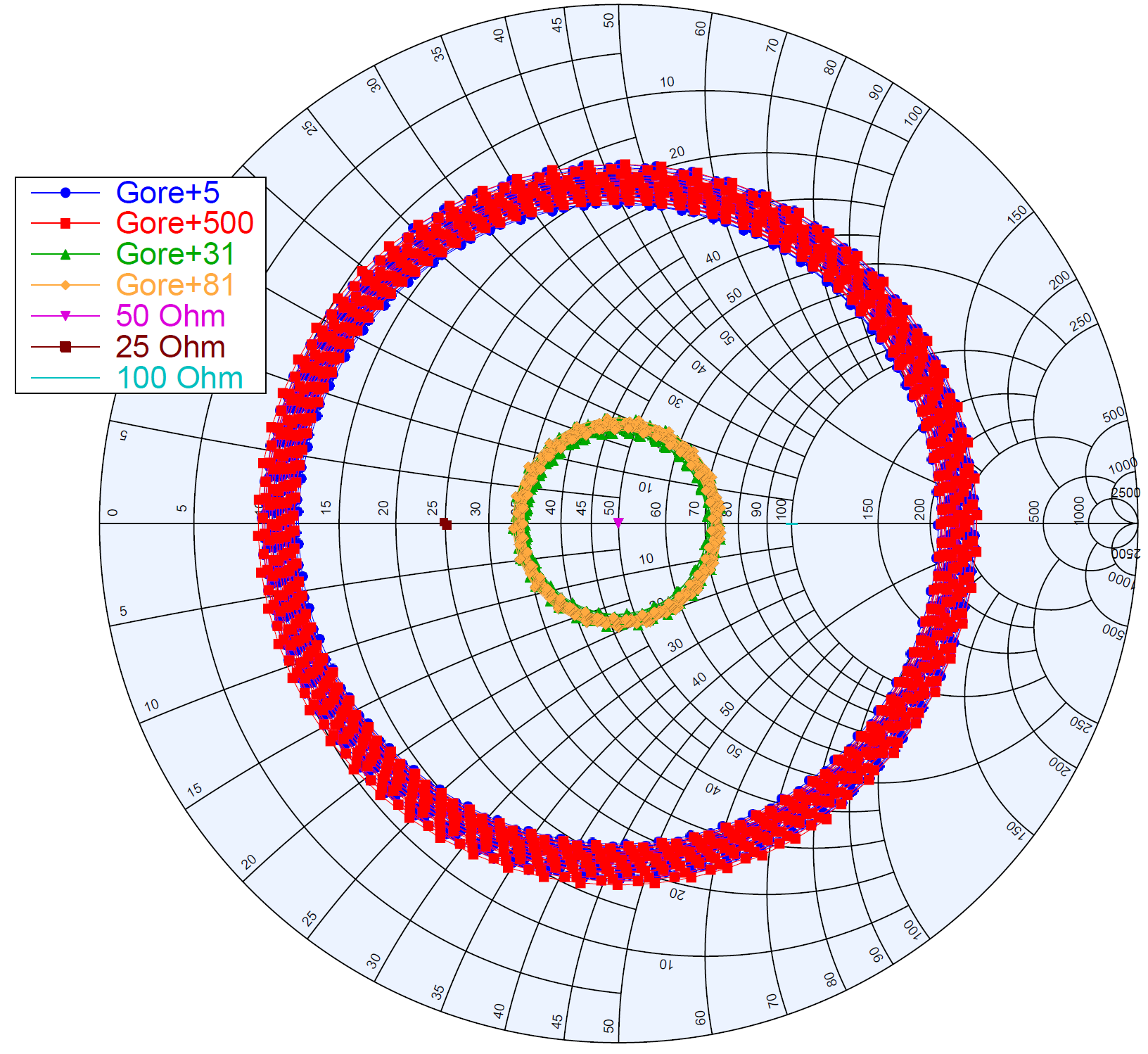}
  \caption{Smith chart (Argand diagram) showing the measured complex impedance of the calibrators used in the Bayesian pipeline across a range of frequencies. 
  \label{f:smith}}
\end{figure}

To verify the performance of our pipeline and highlight features of the algorithm, we evaluate the results of self-consistency checks using empirical models of data based on measurements taken in the laboratory. To make this data as realistic as possible, we used actual measurements of the reflection coefficients of many types of calibrators (see \cref{tab:calibrators}) to generated power spectral densities using \cref{eqn:pant,eqn:pl,eqn:pns} given a set of realistic model noise wave parameters along with some assumptions about the noise, which are described in \cref{chap:solution}. The impedance of the calibrators which were measured with a vector network analyser (VNA) and used in our pipeline are shown on a Smith chart in \cref{f:smith}

\begin{table}
 \caption{Table of calibrators used in the creation of our empirical data models for analysis. Calibrators are added in pairs in the order below when increasing the number of calibration sources used by our algorithm.}
 \label{tab:calibrators}
 \begin{tabular}{lcc}
  \hline
  Calibrator & Temperature\\
  \hline
  Cold load ($50 \ \Omega$) & 298 K\\
  Hot load ($50 \ \Omega$) & 373 K\\
  Gore cable $+ 5 \ \Omega$ & 298 K\\
  Gore cable $+ 500 \ \Omega$ & 298 K\\
  Gore cable $+ 31 \ \Omega$ & 298 K\\
  Gore cable $+ 81 \ \Omega$ & 298 K\\
  $25 \ \Omega$ resistor & 298 K\\
  $100 \ \Omega$ resistor & 298 K\\
  \hline
 \end{tabular}
\end{table}

We start by demonstrating the importance of correlation between noise wave parameters when determining their values to provide a better calibration solution for the reduction of systematic features in the data such as reflections (\cref{chap:correlation}). We then show the increased constraints on these noise wave parameters attributed to the inclusion of more calibrators than the standard number of four (\cref{chap:multCal}). Following this, we illustrate the effectiveness of model selection for the optimisation of individual noise wave parameters to prevent the loss of information resulting from overfitting or underfitting of the data (\cref{chap:opt}). Finally, these features are incorporated into a calibration solution applied to a $50 \ \Omega$ load (\cref{chap:solution}).

\subsection{Correlation between noise wave parameters}\label{chap:correlation}
In this section, we show the first major feature of our Bayesian pipeline; the consideration of correlation between noise wave parameters when deriving their values. This is best demonstrated when noise is introduced in an idealised way as to retain a form matching the Gaussian form of our mathematical model. To do this, empirical models of power spectral densities are calculated from \cref{eqn:pant,eqn:pl,eqn:pns} using measurements of $\G{rec}$, $\Ga$ and $\T{cal}$ for the cold and hot loads, as well as a set of realistic noise wave parameters. Gaussian noise of one unit variation is then added to the $\T{cal}$ measurements after the calculation to conserve its Gaussian form. This data is submitted to our algorithm and the resulting posterior distributions for coefficients of the polynomial noise wave parameters are compared to the initial values.

\begin{figure}
  \centering
  \includegraphics[width=\columnwidth]{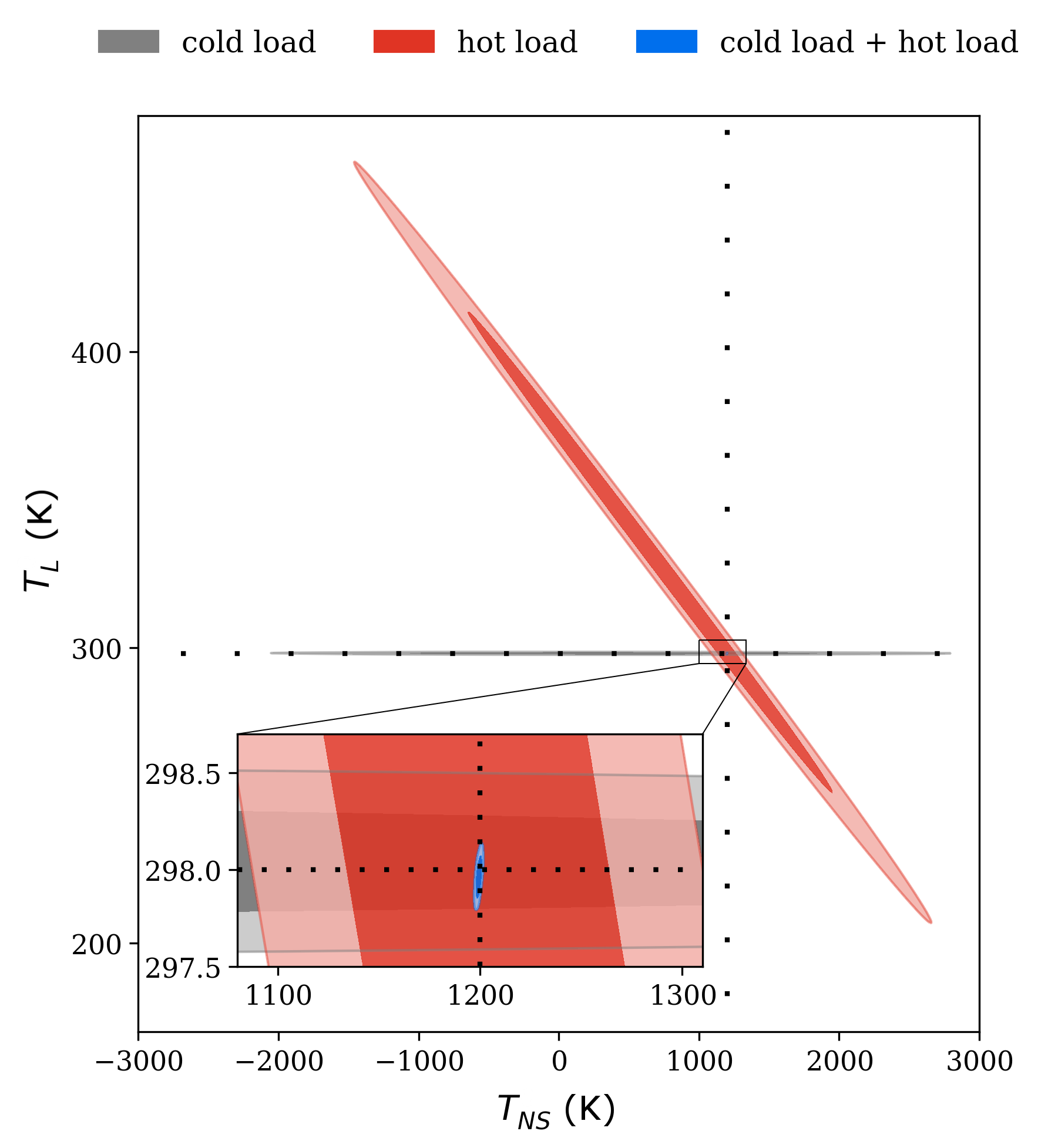}
  \caption{Plot showing the joint posteriors of $\T{L}$ and $\T{NS}$ for models using the cold load, the hot load, and both loads concurrently shown as the grey, red and blue posteriors respectively. The black cross hairs mark the noise wave parameter values used to generate data submitted to the pipeline. A zoom-in of the posterior intersection is provided to illustrate the constraint of noise wave parameter values attributed to the correlation between parameters. \label{f:goodplot}}
\end{figure}

Such posterior distributions can be seen in \cref{f:goodplot} showing the results of models using only the cold load (grey posterior), only the hot load (red posterior) and using both loads in tandem (blue posterior). For these calculations we chose a set of model noise wave parameters as constants across the frequency band;
\begin{align*}
  & \T{unc} = 250 \ \mathrm{K} \\
  & \T{cos} = 190 \ \mathrm{K} \\
  & \T{sin} = 90 \ \mathrm{K} \\
  & \T{NS} = 1200 \ \mathrm{K} \\
  & \T{L} = 298 \ \mathrm{K}
\end{align*}

In \cref{f:goodplot}, a strong correlation between the $\T{L}$ and $\T{NS}$ is evident as the hot-load posterior is highly skewed as expected from \cref{eqn:xl,eqn:xns}. The resulting intersection of posteriors from the individual loads facilitate the derivation of noise wave parameters as the dual-load posterior is found within the region of posterior overlap crossing with the values of the model shown in the inset of \cref{f:goodplot}. Retrieval of the noise wave parameter values using correlations between them found in the data demonstrate the relevance of this information which is not taken into account in previous calibration techniques.

\subsection{Constraints with additional calibrators}\label{chap:multCal}
\begin{figure}
  \centering
  \includegraphics[width=\columnwidth]{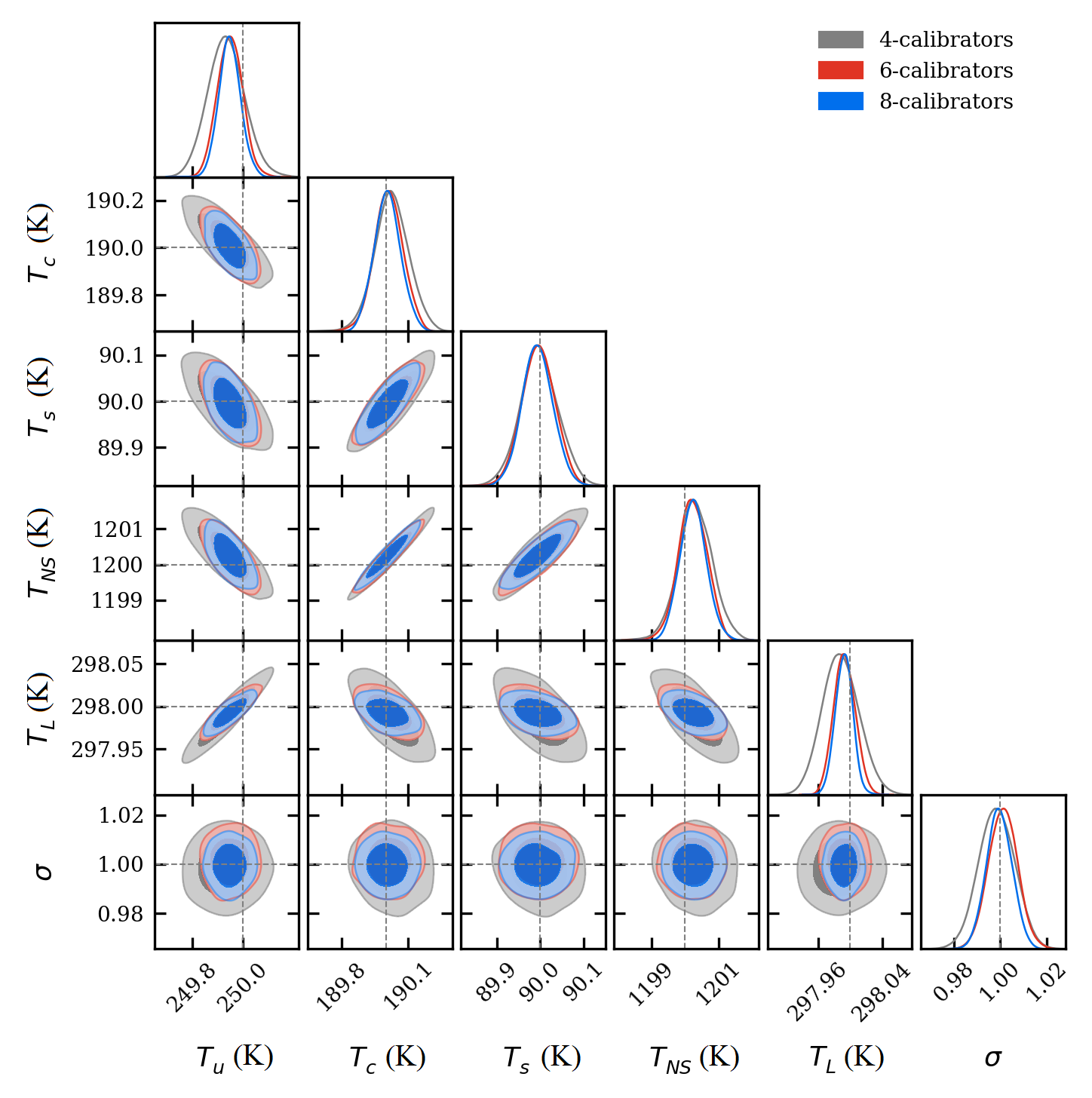}
  \caption{Posterior results of our pipeline using data from four, six and eight calibrators shown in grey, red and blue, respectively. Cross hairs mark the values of noise wave parameters used to generate the data. These values fall within $1\sigma$ of the posterior mean values. We can see that the constraint on noise wave parameter values increases with the number of calibrators used in our pipeline which is encouraging. \label{f:linearall}}
\end{figure}

A nice feature of our pipeline is the ability to include as many calibrators as required to constrain the calibration parameters. For analysis, six more calibrators are introduced in pairs following the order presented in \cref{tab:calibrators}. We include data generated from measurements of multiple resistors terminating a high quality 25 m cable made by GORE\textsuperscript \textregistered.  Data for these calibrators is once again generated using fixed terms and Gaussian noise of one unit variation added to $\T{cal}$ as discussed above. \Cref{f:linearall} shows the results of models using four, six, and eight calibrators. 

As shown, the inclusion of more calibrators increases the constraint on the resulting noise wave parameters. However, we note that after the inclusion of four calibrators, the relative additional constraint decreases with each additional calibrator and thus the use of more than eight calibrators would be unnecessary. The values of noise wave parameters used to generate the data as indicated by the cross hairs in \cref{f:linearall} all fall within $1\sigma$ of our pipeline's resulting posterior averages for models using all eight calibrators.


\subsection{Optimisation of individual noise wave parameters}\label{chap:opt}
The final highlight of our Bayesian pipeline is a the use of machine learning techniques to optimise individual noise wave parameters. This is advantageous as a blanket set of order-seven polynomials applied to all noise wave parameters, such as done in the EDGES experiment, may underfit or overfit individual parameters and misidentify systematics or information about the signal being measured. 

The optimisation procedure compares the evidences (\cref{eqn:ev}) of different models to determine the vector of noise wave parameter polynomial coefficients $\mathbfit{n}$ that best describes the data as briefly mentioned at the end of \cref{chap:bayes}. Since the model favoured by the data will have the highest evidence, we use a steepest descent procedure to compare models in `$\mathbfit{n}$-space' and determine the direction of the gradient in `evidence-space'. After multiple iterations, this brings us to the model with the maximal evidence. Since $\mathbfit{n}$ consists of five numbers corresponding to the number of polynomial coefficients for each of the five noise wave parameters, models are generated by individually increasing each index of $\mathbfit{n}$ by 1. We expect the evidence to follow an `Occam's cliff,' in which the evidence sharply increases preceding the optimal $\mathbfit{n}$ with a slow fall off following the maximum.

\begin{figure}
  \centering
  \includegraphics[width=\columnwidth]{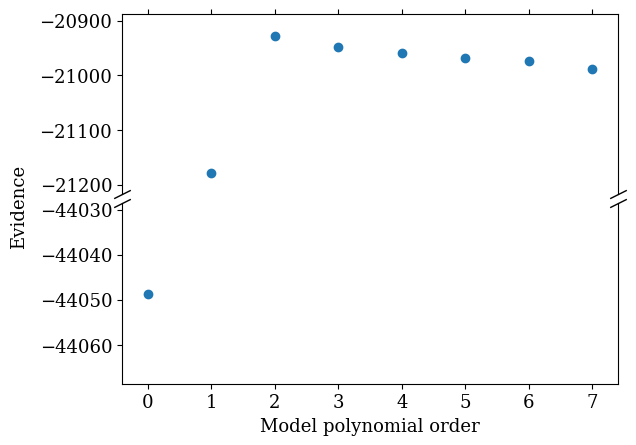}
  \caption{Evidence of multiple models are plotted which display the Occam's cliff. Data is generated using noise wave parameters as order-2 polynomials. We see that for the model with the highest evidence, that is, the model favoured by the data, the number of polynomial coefficients matches that of the model noise wave parameters.\label{f:evidence}}
\end{figure}

To demonstrate this, data is generated using measurements from all eight calibrators of \cref{tab:calibrators} and noise wave parameters as second-order polynomials
\begin{align*}
  & \T{unc} = x^2 -3x + 250 \ \mathrm{K} \\
  & \T{cos} = 2x^2 + 190 \ \mathrm{K} \\
  & \T{sin} = 3x^2 + 8x + 90 \ \mathrm{K} \\
  & \T{NS} = 4x^2 + 5x + 1200 \ \mathrm{K} \\
  & \T{L} = 5x^2 + 10x + 298 \ \mathrm{K} \\
\end{align*}
where $x$ is our normalised frequency. Gaussian noise of one unit variation is applied to the calibrator input temperatures as before. The evidences of various models are plotted in \cref{f:evidence} in which an Occam's cliff can be seen peaking at polynomial order two. As expected from the plot, the steepest descent algorithm finds that noise wave parameters modelled as second-order polynomials best describe the data.

\subsection{Application with realistic noise}\label{chap:solution}
\begin{figure}
  \centering
  \includegraphics[width=\columnwidth]{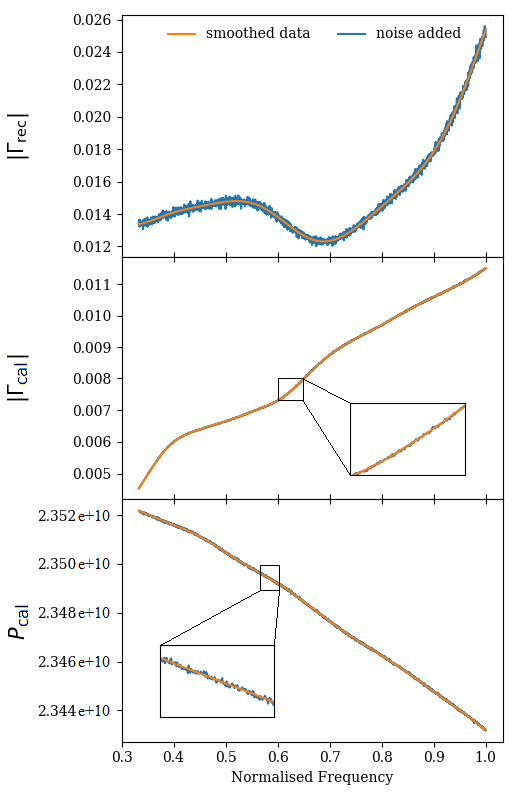}
  \caption{Power spectral densities and reflection coefficients for the receiver and the cold load generated under our realistic noise model. \label{f:calQualities}}
\end{figure}

To demonstrate the robustness of our pipeline, we conducted self-consistency checks using empirically modelled data with a more complicated noise model. This data was generated using reflection coefficients of eight calibrators and the receiver measured in the laboratory. These reflection coefficients were then smoothed using a cubic smoothing spline \citep{spline} in order to maintain their approximate shape over frequency. The same second-order noise wave parameters detailed in \cref{chap:opt} are used with the reflection coefficients to generate our model power spectral densities. Following this, we added of order 1\% Gaussian noise independently to the smoothed $\G{rec}$ and $\Ga$ as well as $\psd{cal}$ to more accurately represent the instrument noise from measurement equipment such as vector network analysers. No noise was added to the calibrator input temperatures. This results in a model that does not match the Gaussian form of our mathematical model as in the previous sections and thus does not demonstrate the features of our pipeline as explicitly, but is more representative of data set expected from measurements in the field. Data for the receiver and the cold load generated using this noise model are shown in \cref{f:calQualities}.

\begin{figure*}
  \centering
  \includegraphics[width=.9\textwidth]{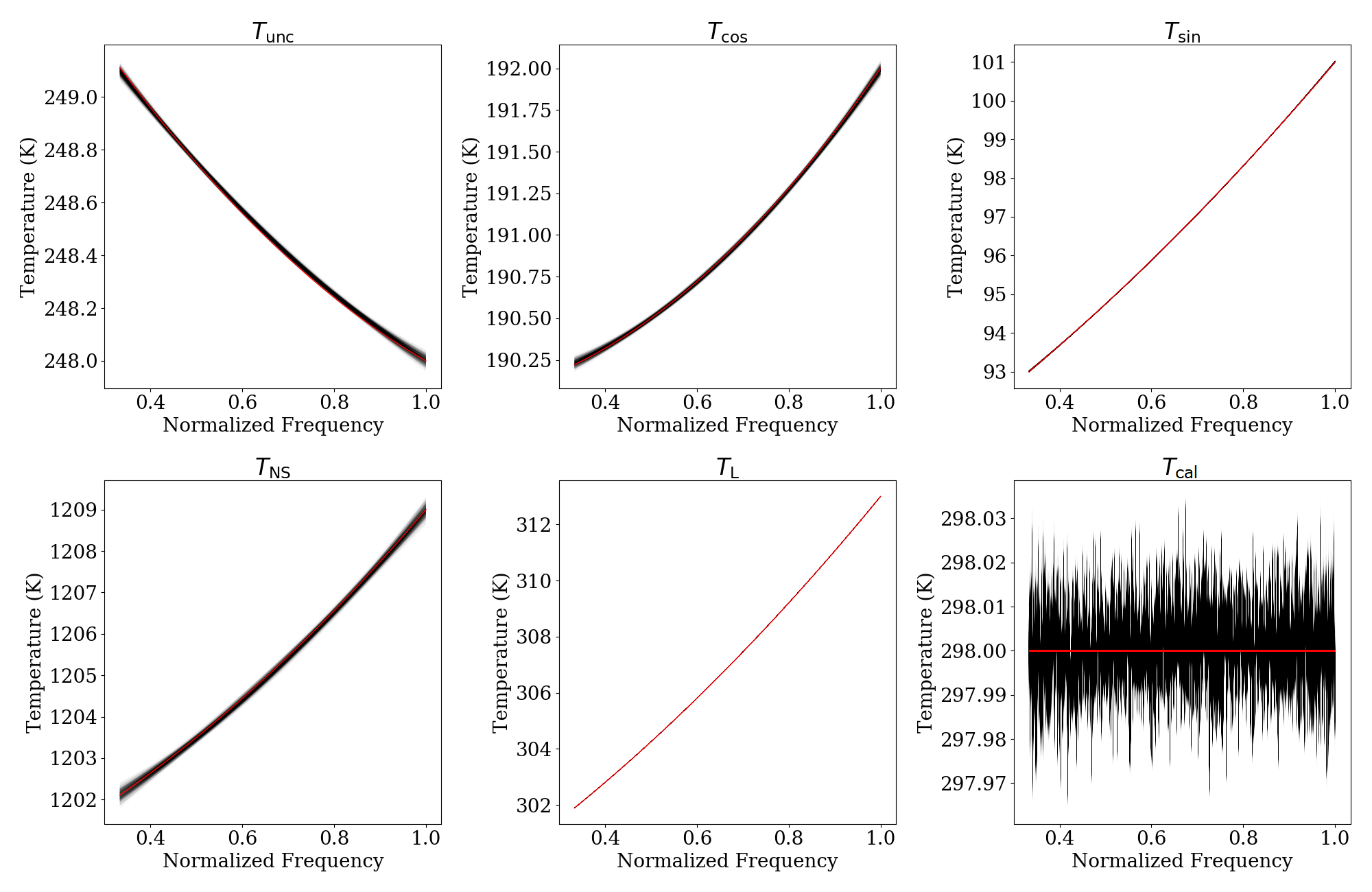}
  \caption{Results from 1000 samples using data generated with our more realistic noise model (shown in black). The second-order noise wave parameters shown in red are used to generate the data inputted to our pipeline. The polynomial order and values of the noise wave parameters that best suit the data according to our algorithm match that of the empirical model. This solution is applied to an ambient-temperature load, shown in the bottom right panel as our predictive $\hat{y}$ from \cref{predictive}, and calibrates it to within $1\sigma$ of ambient temperature. \label{f:fgxSamples}}
\end{figure*}

Using data generated for all eight calibrators with our realistic noise model, the calibration algorithm selects optimal polynomial orders matching those of the model noise wave parameters whose values fall within within $1\sigma$ of the posterior peak values as shown in \cref{f:fgxSamples}. For these higher order tests, we use fgivenx plots which condense noise wave parameter posteriors into samples that can be compared to the model parameter values instead of comparing each individual coefficient \citep{fgx}. 

When this calibration model is used to calibrate an ambient-temperature $50 \ \Omega$ load, the RMS error between the calibrated temperature and the measured temperature is 8 mK, well within the $1\sigma$ noise level (bottom right panel of \cref{f:fgxSamples}). This level of accuracy is comparable to the 26 mK noise floor estimated for the EDGES pipeline in 2016 \citep{calpap}. 

By individually adjusting each component of noise arising in our realistic noise model, we may determine what kind of noise our calibration algorithm is most sensitive to, as well as calculate the maximum amount of noise permissible for a specified level of systematic feature reduction. These topics are intended to be explored in a future work.

\section{Conclusions} \label{conclusions} 
Here we presented the development of a calibration methodology based on the procedure used by EDGES but with key improvements to characterise reflections arising at connections within the receiver. Our pipeline utilises the Dicke switching technique and a Bayesian framework in order to individually optimise calibration parameters while identifying correlations between them using a dynamic algorithm to be applied in the same environment as the data acquisition. In a comprehensive investigation, we have evaluated our algorithm's interpretation of empirical models of data which have been generated from known noise wave parameters and a realistic noise model. The solution, applied to an ambient-temperature $50 \ \Omega$ load, produces a calibrated temperature with an RMS residual temperature of 8 mK. Future work for the pipeline regards application of real calibrator data, optimisation of noise wave parameter coefficients through marginalisation techniques and incorporation into an end-to-end simulation based on an entire experimental apparatus to better understand error tolerances. The flexibility of the algorithm attributed to our novel approach allows its application to any experiment relying on similar forms of calibration such as REACH \citep{reach}, were we intend to use this for in-the-field and on-the-fly radiometer calibration.


\section*{Acknowledgements}


ILVR would like to thank S. M. Masur for her helpful comments. WJH was supported by a Gonville \& Caius Research Fellowship, STFC grant number ST/T001054/1 and a Royal Society University Research Fellowship. We would like to thank the The Cambridge-Africa ALBORADA Research Fund for their support. We would also like to thank the Kavli Foundation for their support of the REACH experiment.

\section*{Data Availability}


The data underlying this article will be shared on reasonable request to the corresponding author.



\bibliographystyle{mnras}
\bibliography{references} 






\bsp	
\label{lastpage}
\end{document}